%% file: main.tex
\documentclass[sigconf,nonacm]{acmart}

\AtBeginDocument{%
  }

\setcopyright{acmlicensed}
\copyrightyear{2018}
\acmYear{2018}
\acmDOI{XXXXXXX.XXXXXXX}
\acmConference[Conference acronym 'XX]{Make sure to enter the correct
  conference title from your rights confirmation email}{June 03--05,
  2018}{Woodstock, NY}

\makeatletter

\acmISBN{978-1-4503-XXXX-X/2018/06}

\usepackage{algorithm}
\usepackage{algpseudocode}
\usepackage{cancel}
\usepackage{amsmath,amssymb}
\usepackage{mathtools}
\usepackage{booktabs}
\usepackage[table]{xcolor}
\usepackage{enumitem}
\usepackage{float}
\usepackage{wrapfig,lipsum}
\usepackage{comment}
\usepackage{pifont} 
\microtypesetup{expansion=false}
\theoremstyle{definition}

\theoremstyle{remark}

\usepackage{listings}
\usepackage{booktabs}
\usepackage{amsmath,amssymb}
\usepackage{flafter}

\usepackage{tcolorbox}
\tcbuselibrary{breakable,skins}

\newtcolorbox{examplebox}{
  colback=blue!10,
  colframe=blue!20,
  arc=2mm,
  fonttitle=\bfseries,
  boxrule=0mm,
  boxsep=1mm,
  left=0mm,
  right=0mm,
  top=0mm,
  bottom=0mm
}

\newtcolorbox{pbox}{
  colback=black!5,
  colframe=black!30,
  arc=2mm,
  fonttitle=\bfseries,
  boxrule=0mm,
  boxsep=1mm,
  left=0mm,
  right=0mm,
  top=0mm,
  bottom=0mm
}

\newtcolorbox{defbox}{
  colback=white,
  colframe=black!100,
  arc=1mm,
  fonttitle=\bfseries,
  sharp corners=all,
  boxrule=0.2mm,
  boxsep=1mm,
  left=0mm,
  right=0mm,
  top=0mm,
  bottom=0mm
}

\definecolor{promptwrap}{gray}{0.6}
\lstdefinestyle{promptstyle}{
  basicstyle=\ttfamily\scriptsize,
  breaklines=true,
  breakatwhitespace=false,
  breakindent=0pt,
  columns=fixed,
  keepspaces=true,
  showstringspaces=false,
  upquote=true,
  frame=none,
  xleftmargin=0pt,
  xrightmargin=0pt,
  aboveskip=2pt,
  belowskip=0pt,
  postbreak=\mbox{\textcolor{promptwrap}{$\hookrightarrow$}\space},
}

\newtcolorbox{promptbox}{
  breakable,
  enhanced,
  pad at break*=0.5mm,
  lines before break=1,
  before upper={\clubpenalty=0 \widowpenalty=0 \interlinepenalty=0 \brokenpenalty=0},
  colback=black!3,
  colframe=black!35,
  arc=1mm,
  boxrule=0.25mm,
  boxsep=1mm,
  left=1mm,
  right=1mm,
  top=1mm,
  bottom=1mm
}

\newtheoremstyle{promptthm}
  {2pt}          
  {0pt}          
  {\normalfont}  
  {0pt}          
  {\bfseries}    
  {.}            
  {\newline}     
  {}             
\theoremstyle{promptthm}
\newtheorem{prompt}{Prompt}

\begin{document}
\title{A Vision for a Logic-Based Workflow Formulation Framework and an Agentic Execution Pipeline for Pattern Engineering}

\author{Alankrit Moses}
\affiliation{%
  \institution{University Of Texas At Arlington}
  \country{Texas, USA}
}
\email{axm1006@mavs.uta.edu}

\author{Raghav Mittal}
\affiliation{%
  \institution{University Of Texas At Arlington}
  \country{Texas, USA}
}
\email{rxm0006@mavs.uta.edu}

\author{Gautam Das}
\affiliation{%
  \institution{University Of Texas At Arlington}
  \country{Texas, USA}
}
\email{gdas@uta.edu}

\author{Chengkai Li}
\affiliation{%
  \institution{University Of Texas At Arlington}
  \country{Texas, USA}
}
\email{cli@uta.edu}

\input{sections/sec-abstract}

\maketitle

\input{sections/sec-intro}

\input{sections/sec-related}
\input{sections/sec-framework}
\input{sections/sec-method}
\input{sections/sec-eval}
\input{sections/sec-conclusion}

\bibliographystyle{ACM-Reference-Format}
\bibliography{references.bib}
\input{sections/prompts}

\end{document}

%% file: sections/sec-abstract.tex
\begin{abstract}
Organizations increasingly depend on insights drawn from data to inform consequential decisions. Many concern structural behaviors, which we call patterns. Producing them may require working with multiple pattern types, performing different operations over them, and composing those operations into workflows, an activity we call pattern engineering. No principled framework exists for organizing such workflows: analysts devise each by hand, and existing systems automate particular analytical objectives rather than representing pattern operations in a general, composable form. We introduce a typed framework that represents each pattern through a fixed set of modular components, and expresses pattern operations as queries which specify input and output components. Chaining queries composes workflows, while execution of any individual query remains unconstrained. We instantiate the framework with an LLM-based agentic executor and evaluate example-mining, example-alignment, and pattern-induction queries over two temporal patterns, using World Intellectual Property Organization (WIPO) reports and patent data. All three queries are carried out across four LLM settings, which perform differently by target component, while the traces allow us to localize errors by execution stage. 
\end{abstract}

%% file: sections/sec-intro.tex
\section{Introduction}\label{sec:intro}
Organizations increasingly depend on insights drawn from data to inform consequential decisions. These insights reveal how a system behaves, what is changing within it, and where risks or opportunities may be emerging. The World Intellectual Property Organization (WIPO)'s patent landscape report~\footnote{\url{https://tind.wipo.int/record/28192}}
on Ritonavir, an antiretroviral drug used to treat HIV, presents insights of this kind. Its analysis is based on a publicly released tabular dataset~\footnote{\url{https://tind.wipo.int/record/50100}}
of 805 patent families, described by 73 attributes such as year, country, etc.; Figure~\ref{fig:pattern-instance}(a) shows a truncated sample. An independent WIPO evaluation later found that the report had been used to identify possible licensing arrangements for the drug;~\footnote{\url{https://www.wipo.int/meetings/en/doc_details.jsp?doc_id=217682}} licenses covering Ritonavir-based formulations followed, reaching at least 102 low- and middle-income countries.~\footnote{\url{https://medicinespatentpool.org/licence-post/lopinavir-ritonavir-lpv-r}}

Aggregating the Ritonavir dataset in Figure~\ref{fig:pattern-instance}(a) by year yields the annual series in Figure~\ref{fig:pattern-instance}(c), where each value represents the number of patent families with filings in that year. The time series of counts shows a steady increase to a peak in 1999, followed by a second rise culminating in 2006 and a subsequent decline. This behavior exemplifies a \emph{trajectory shift}, a common structural pattern in which a tracked quantity changes direction after a turning point. Similar structural forms recur across different data windows and domains. We use the term \emph{pattern} to refer to such abstract structures in data.

\newcommand{\panel}[1]{\raisebox{\height}{#1}}

\newlength{\lblshift}\setlength{\lblshift}{0pt} 
\newcommand{\plabel}[1]{\raisebox{\dimexpr\height-\lblshift}{\textbf{#1}}}

\begin{figure*}[t]
\centering
\small
\setlength{\fboxsep}{3pt}
\makebox[\textwidth][c]{%
\begin{tabular}{@{}c@{\hspace{0.5em}}c@{\hspace{1.2em}}c@{\hspace{0.5em}}c@{\hspace{1.2em}}c@{\hspace{0.5em}}c@{}}
\plabel{(a)} &
\panel{%
\scriptsize
\renewcommand{\arraystretch}{1}
\begin{tabular}{@{}lcc@{}}
\toprule
\textbf{Patent family} & \textbf{Year} & $\cdots$ \\
\midrule
WO2001021603 & 1999 & $\cdots$ \\
US6232333    & 1996 & $\cdots$ \\
CN101613325  & 2008 & $\cdots$ \\
$\cdots$     & $\cdots$ & $\cdots$ \\
\bottomrule
\end{tabular}}
&
\plabel{(b)} &
\panel{\fbox{\parbox{0.48\textwidth}{%
\footnotesize
{\scriptsize \underline{Excerpt, Sec.~3.2 of the Ritonavir report
:}}
Figure 1 below describes patenting activity over time. The graph presents the number of patent families (Y axis) over the earliest filing (priority) year (X axis). \dots{} The graph shows a large increase in filings up to 1999, and then a second spike up to 2006. The reduction after 2006 may be attributed to typical lags between a first earliest priority filing \dots{} and the first published application.}}}
&
\plabel{(c)} &
\panel{%
\scriptsize
\setlength{\tabcolsep}{0.5pt}
\renewcommand{\arraystretch}{1}
\begin{tabular}{@{}l*{7}{c}@{}}
\toprule
Year     & 1996 & 1997 & 1998 & 1999 & 2000 & 2001 & 2002 \\
Families & 20 & 25 & 43 & 55 & 36 & 49 & 63 \\
\midrule
Year     & 2003 & 2004 & 2005 & 2006 & 2007 & 2008 & 2009 \\
Families & 64 & 74 & 80 & 94 & 85 & 65 & 29 \\
\bottomrule
\end{tabular}}
\end{tabular}%
}
\caption{(a) Truncated view of the Ritonavir patent dataset, with attribute renamed for ease of exposition. (b) Report excerpt on patenting activity over time. (c) Annual patent family counts by earliest priority year; 23 pre-1996 families omitted.}
\label{fig:pattern-instance}
\end{figure*}

Analysts rarely work with a single pattern when drawing insights from data. WIPO's patent landscape report on generative AI,~\footnote{\url{https://www.wipo.int/web-publications/patent-landscape-report-generative-artificial-intelligence-genai/en/index.html}}
for instance, identifies both a trajectory shift pattern (Figure~\ref{fig:pattern-cards}(a3)) and a \emph{growth divergence} pattern, in which two quantities move steadily further apart over an interval (Figure~\ref{fig:pattern-cards}(b3)). Analysts must also perform different operations over these patterns. In the study that introduced the El Ni\~no Modoki phenomenon~\cite{ashok2007elninomodoki}, researchers formulated a mathematical definition capturing an unusual recurring behavior they had observed (\emph{pattern induction}), then applied it to historical data to find further occurrences (\emph{example mining}); the output of the pattern induction operation became the input to the example mining operation, composing their study's analytical workflow. Other operations are equally natural, such as translating between a definition and its natural-language interpretation (\emph{pattern translation}), or proposing further patterns from ones already known (\emph{pattern suggestion}). This combined activity of representing different patterns, specifying operations over them, and composing those operations into workflows is what we call \textbf{pattern engineering}. 

Pattern engineering, however, remains an ad-hoc craft rather than a systematic discipline, as it lacks a principled framework for structuring and guiding its practice.  The design of a study is left largely to the individual analyst, who must determine which patterns are relevant, which pattern operations to apply, and how to order and connect them into an appropriate workflow. Because that workflow is built around the particular patterns and procedures it happens to use, little of it carries over to the next study, and introducing a new pattern or operation into it may require substantial redesign. Repeating this open-ended process for every study makes pattern engineering time-consuming and effort-intensive.

Organizations increasingly derive insights from heterogeneous data---including reports, tables, and other modalities---collected across their data lakes. 
We envision pattern engineering as a structured analytical layer over these existing data assets. By explicitly representing data, patterns, and operations over patterns, this layer would connect modalities and enable workflows to be specified and composed independently of any particular pattern type, dataset, document collection, analytical system, or predefined pipeline. A study could thus be defined in terms of what each step requires and produces, rather than the specific machinery used to execute it.

We realize this vision for pattern engineering through a declarative framework inspired by \textsc{Datalog}~\cite{ceri1989datalog}
. The framework represents data and patterns through components. Data has two components corresponding to its textual (\texttt{D.text}) and tabular (\texttt{D.tab}) modalities, while each pattern has three components corresponding to its mathematical definition (\texttt{M}), textual description (\texttt{T}), and a set of examples (\texttt{E}). These components form data and pattern tuples that are used to construct queries. The tuples have fixed component positions inspired by the argument structure of atoms in \textsc{Datalog}. A query assigns roles to these positions by indicating which components are provided as inputs, requested as outputs, or left anonymous. Following \textsc{Datalog}, the notation uses \texttt{\_} for anonymous component positions and extends it with the output marker $?$ and the cardinality annotation \texttt{[n]} to identify a target component and the number of requested results. A query therefore specifies the relationship between available input components and desired output components without prescribing how specifically that relationship should be computed. For example, Q1 in Table~\ref{tab:problem-catalog}, written as $\mathtt{D:(D.text,\_), P:(M,\_,\{(?,\_)[3]})\}$, provides textual data (for instance, an analyst's document corpus) and a mathematical definition (see Figure~\ref{fig:pattern-cards}) as inputs and requests three textual examples of the pattern, while leaving the remaining components anonymous. 

The queries represented by the framework must be resolved by an executor. Since the framework can express mathematically diverse patterns and multiple operations, the executor would naturally require to be general enough to handle mathematical reasoning, natural-language understanding, tabular analysis, and workflow planning. Given the capabilities of LLMs in these domains they are a strong fit to execute these queries. We instantiate the framework through an LLM-based agentic pipeline that comprises five agents---three component specialists, namely the \texttt{M}-Inducer, \texttt{T}-Inducer, and \texttt{e.text-finder}; the orchestrator agent that coordinates the actions of these three; and the \texttt{e.text-selector} agent which, accessible only to the \texttt{e.text-finder}, ranks retrieved candidate passages in selecting textual examples.

For evaluating the preliminary framework design and executor implementation in this exploratory study, we select three relatively simple queries that cover distinct operations while not using \texttt{e.tab} as an output. This choice allows the executor to function without implementing a corresponding \texttt{e.tab}-specialist. A full-fledged executor could implement such an agent in several ways. For example, it could analyze tabular data directly or use an auxiliary SQL or coding agent to retrieve and compute the aggregations needed for tabular-example analysis.\newline
\indent The evaluation is likewise limited only to time-series data from WIPO patent landscape reports and their corresponding tabular patent data. Within this setting, we study two mathematically distinct temporal patterns, \emph{growth divergence} and \emph{trajectory shift}, under three representative queries from Table~\ref{tab:problem-catalog}. These involve mining textual examples from a supplied mathematical definition (Q1), recovering textual evidence corresponding to a tabular instance (Q2), and inducing a definition and description from paired examples (Q3).

Though not a full-scale evaluation, the setup helps us probe the capacity of the executor and the results provide proof-of-concept evidence that the framework can express, and the executor can carry out, multiple pattern-engineering tasks across LLM backends. The evaluated backends rank differently depending on whether the target is \texttt{e.text} or the induced \texttt{M} and \texttt{T} components, suggesting that the framework can also serve as a diagnostic tool for model--component affinity. The execution traces further show the orchestrator introducing intermediate components that are not requested as final outputs, using them to compose multi-step executions, and recording the stages at which errors arise.

Future work will extend the framework beyond textual and tabular components to visual and other structured modalities. It will also require a code- or SQL-backed \texttt{e.tab}-specialist, experiments assigning different models and reasoning budgets to individual component agents, and evaluation over larger pattern libraries, corpora, and domains with broader gold annotations and human assessment.

The datasets and code used for this study, as well as the prompts that instantiate the executor and its communication protocol used in our experiments can be found in supplementary files as well as our code repository, available at \url{https://github.com/idirlab/patternengineering}.

%% file: sections/sec-related.tex
\begin{figure}[t]
  \centering
\includegraphics[width=0.48\textwidth]{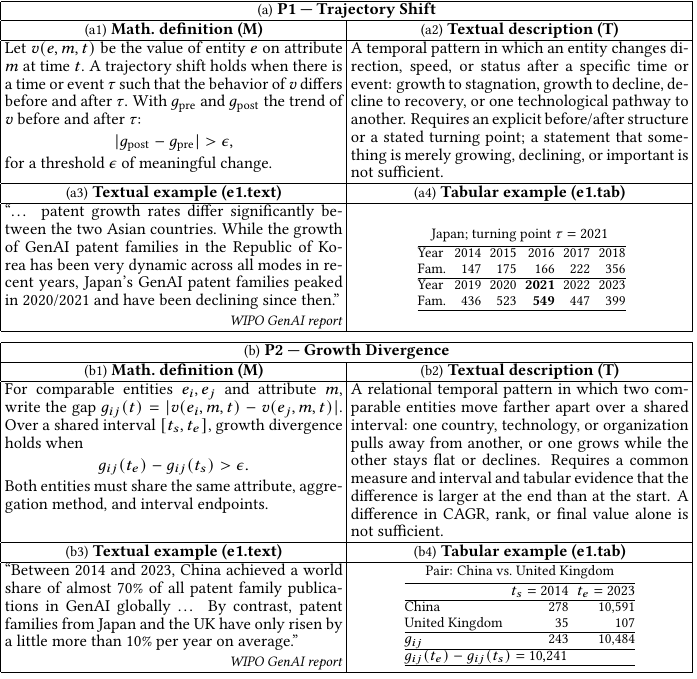}
 \caption{Pattern cards for trajectory shift and growth divergence, with each pattern's math. definition (\texttt{M}), text description (\texttt{T}), and paired text + tabular examples (\texttt{e.text}, \texttt{e.tab}).}
  \label{fig:pattern-cards}
\end{figure}
\section{Related Work}
\label{sec:related}

\textbf{Pattern mining.}\hspace{2mm} A large body of prior work focuses on discovering instances of predefined patterns from data. Classical examples include association-rule and frequent-pattern mining~\cite{han2007frequent}, subgroup discovery~\cite{atzmueller2015subgroup}, contrast patterns~\cite{dong1999emerging}, as well as efforts toward spatio-temporal pattern discovery~\cite{atluri2018spatio}. Fact-finding systems likewise fix a notion of an interesting fact and search for instances satisfying it~\cite{factwatcher}.  These works map to the example-mining setting represented by Q4, where \texttt{M} is supplied and matching instances are recovered from tabular data. Our contribution is not a new mining algorithm, but a framework in which such tasks, together with other pattern-engineering tasks, are represented as queries, and an executor that resolves queries from pattern components rather than fixed solutions for specific pattern types.

\textbf{Pattern representation systems.}\hspace{2mm} Automated insight-discovery systems provide more unified representations of patterns. Systems such as \textsc{QuickInsights}~\cite{ding2019quickinsights} and \textsc{MetaInsight}~\cite{ma2021metainsight} organize analysis around predefined insight types whose detection, scoring, and type-specific properties are implemented by system developers, while \textsc{InsightPilot}~\cite{ma2023insightpilot} builds LLM-driven exploration over such underlying insight engines. Pattern-based management systems (PBMSs) similarly treat patterns as first-class objects represented through system-defined schemas, measures, and formulas, and provide languages for querying pattern bases~\cite{terrovitis2007patternbases}. 

In all of these, what counts as an admissible pattern is decided when the system is built, so extending to a new pattern means extending the system at development time. Our framework instead fixes only the component structure, while what constitutes a pattern is supplied through its \texttt{M} and \texttt{T} components. 
It likewise does not take a fixed inventory of operations as primitives. New pattern definitions and pattern-engineering tasks can therefore be expressed without registering a new pattern type or operation in advance.

\textbf{Declarative and LLM foundations.}\hspace{2mm} Our queries follow the declarative structure of \textsc{Datalog}, in which relations are specified by rules over fixed argument positions without prescribing an execution procedure~\cite{ceri1989datalog}
. Recent LLM-based data systems expose declarative interfaces over rows and documents~\cite{patel2025semanticoperators}, structure the model interaction itself~\cite{khattab2024dspy}, or coordinate specialized agent roles~\cite{wu2024autogen}. While these motivate our agentic executor, we note that each of these fix their operators or agent topologies in advance, whereas ours follows from the component types the query names.

%% file: sections/sec-framework.tex
\section{Pattern Engineering Framework}
\label{sec:framework}

The framework is inspired by the declarative and positional structure of \textsc{Datalog}~\cite{ceri1989datalog}, where components occupy fixed positions within tuples, while queries specify relationships between supplied and requested components without prescribing how they should be resolved. Our notation extends this structure with markers for requested outputs and output cardinality.

A \emph{query} specifies which data and pattern components are available and which should be produced, without prescribing how the task is executed. 
We group queries according to their semantic meaning and refer to each such group as a type of \emph{operation}.  Because operations emerge from the semantic grouping of valid queries rather than being predefined, the framework is not restricted to a closed catalog of operations. Table~\ref{tab:problem-catalog} presents several representative cases. Q1 and Q4 are structurally different but both perform example mining. Along similar lines, Q3 and Q7 both request a mathematical definition, yet Q3 performs pattern induction because it uses examples to derive the definition whereas Q7 performs pattern formalization as it formalizes a definition for an existing text description. Further observe how Q2 is distinguished as example alignment because it recovers the textual counterpart of a supplied tabular example rather than mine examples from a supplied pattern definition.

Queries can also be chained into workflows by supplying a component produced by one query as input to another. For example, Q3 may first produce \texttt{M} and \texttt{T} from examples, after which the resulting \texttt{M} can be supplied to Q1 to mine additional textual examples (for instance, recall the induction-then-mining composition of the El Ni\~no Modoki study from Section~\ref{sec:intro}). This allows more complex workflows to be composed while keeping intermediate components available for inspection and reuse.

Each query contains one data tuple \texttt{D} and one or more pattern tuples \texttt{P1}--\texttt{Pn}. The data tuple represents two modalities. Tabular data, denoted by \texttt{D.tab}, refers to structured datasets such as spreadsheets or relational tables used in an analyst's study. For simplicity, in this work \texttt{D.tab} refers to a single unified table. Textual data, denoted by \texttt{D.text}, corresponds  to a set of textual reports, which refer to information expressed in natural language, such as patent landscape reports, corporate reports, or technical documents.

\begin{table}[t]
\centering
\scriptsize
\caption{Representative typed pattern queries expressible in the framework. Q1--Q3 are evaluated in our case study.}
\label{tab:problem-catalog}
\begin{tabular}{@{}>{\centering\arraybackslash}m{1.5em}@{\hspace{2pt}}|p{0.71\columnwidth}@{\hspace{1.5pt}}|m{0.21\columnwidth}@{}}
\toprule
\textbf{ID} & \textbf{Typed Query and Description} & \textbf{Operation} \\
\midrule
Q1 & $\mathtt{D:(D.text,\_)} \qquad\mathtt{P:(M,\_,\{(?,\_)[3]\})}$\newline
\textbf{Description:}~Given reports and a mathematical definition, recover three textual examples that instantiate the pattern. & Example Mining \\
\midrule
Q2 & $\mathtt{D:(D.text,\_)} \qquad\mathtt{P:(\_,\_,\{(?,e.tab)\})}$\newline
\textbf{Description:}~Given reports and a tabular example, recover the corresponding textual example. & Example Alignment \\
\midrule
Q3 & $\mathtt{D:(\_,\_)} \qquad\mathtt{P:(?,?,\{(e_1.text,e_1.tab),(e_2.text,e_2.tab),}$
\newline$\mathtt{(e_3.text,e_3.tab),(e_4.text,e_4.tab)\})}$\newline
\textbf{Description:}~Given four paired examples, induce the mathematical definition and textual description of the pattern. & Pattern Induction \\
\midrule
Q4 & $\mathtt{D:(\_,D.tab)} \qquad\mathtt{P:(M,\_,\{(\_,?)[n]\})}$\newline
\textbf{Description:}~Given tabular data and a mathematical definition, recover $n$ tabular examples satisfying the pattern. & Example Mining \\
\midrule
Q5 & $\mathtt{D:(D.text,\_)} \qquad\mathtt{P:(M,?,\_)}$\newline
\textbf{Description:}~Given reports and a math definition, produce a text description that interprets the pattern in the context of the reports. & \mbox{Pattern Interpretation} \\
\midrule
Q6 & $\mathtt{D:(D.text,D.tab)} \qquad\mathtt{P:(\_,\_,\{(e_1.text,e_1.tab),(?,?)[3]\})}$\newline
\textbf{Description:}~Given one paired example, recover three additional paired examples of the same pattern. & Example Mining \\
\midrule
Q7 & $\mathtt{D:(\_,\_)} \qquad\mathtt{P:(?,T,\{(e.text,e.tab)\})}$\newline
\textbf{Description:}~Given a textual description and an example pair, synthesize a mathematical definition consistent with both. & \mbox{Pattern Formalization} \\
\midrule
Q8 & $\mathtt{D:(\_,\_)} \qquad\mathtt{P1:(M,T,\_)} \qquad\mathtt{P2:(?,?,\_)}$\newline
\textbf{Description:}~Given the mathematical definition and textual description of an existing pattern, suggest a new pattern by producing its mathematical definition and textual description. & Pattern Suggestion \\
\bottomrule
\end{tabular}
\end{table}

Each pattern tuple represents one pattern through three components: a mathematical definition (\texttt{M}), a textual description (\texttt{T}), and a set of examples (\texttt{E}). The mathematical definition states the formal conditions under which the pattern occurs, as shown for trajectory shift and growth divergence in Figures~\ref{fig:pattern-cards}(a1) and~\ref{fig:pattern-cards}(b1). Thus, a pattern can provide its own formal criterion for what constitutes an instance, rather than being restricted to a predefined form. 

The textual description (\texttt{T}) provides a natural-language interpretation of the pattern, capturing its meaning, scope, and qualifying conditions. It complements (\texttt{M}) by expressing the pattern in a form suited to unstructured data (text in this case) where an instance may convey the relevant behavior without explicitly stating every mathematical variable or measurement. For example, the growth-divergence passage in Figure~\ref{fig:pattern-cards}(b3) describes China pulling away from the United Kingdom, consistent with \texttt{T} in panel (b2), even though it does not explicitly provide all the quantities used by \texttt{M} in panel (b1).

The examples of a pattern, denoted by \texttt{E}$=\{e_1,e_2,\dots\}$, form a set of pairs. Each pair has two parts, a textual example (\texttt{e.text}) and a tabular example (\texttt{e.tab}). The textual example is an instance of the pattern observed in text data, such as the growth-divergence passage in Figure~\ref{fig:pattern-cards}(b3). The corresponding tabular example is \texttt{e1.tab} in Figure~\ref{fig:pattern-cards}(b4), which gives the endpoint counts for the two countries and the widening gap between them.

The data and pattern tuples assign fixed positions to their components, following the templates \mbox{\texttt{D:(D.text,D.tab)}} for data and \mbox{\texttt{P:(M,T,\{(e.text,e.tab)\})}} for patterns. A named component in its designated position indicates that it is supplied as input. For example, Q1 names \texttt{D.text} and \texttt{M}, indicating that textual data and a mathematical definition are supplied. A question mark ($?$) may occupy a pattern-component position to indicate a requested output. Thus, the \texttt{?} in the \texttt{e.text} position of Q1 requests textual examples, while the two question marks in Q3 request \texttt{M} and \texttt{T}. Only components of pattern tuples can be requested as outputs, and a valid query must contain at least one pattern tuple with at least one component marked \texttt{?}.

An underscore ($\_$) marks a component that is neither supplied as input nor requested as output; in Q1, this applies to \texttt{D.tab}, \texttt{T}, and \texttt{e.tab}. Such a component lies outside the query's input-output specification, although an executor may still derive or use it as an intermediate result. A pattern tuple of the form \texttt{P}$:(\_,\_,\_)$ specifies neither input nor output components and therefore does not participate in the query, so it should be omitted. On the other hand, \texttt{D}$:(\_,\_)$ is valid when the task requires no external data, as in Q3, where the supplied examples alone are used to produce \texttt{M} and \texttt{T}. 
Finally, \texttt{[n]}, which is a repetition quantifier borrowed from regular expressions, specifies the number of examples requested from a target position. Q1 uses \texttt{[3]} to request three textual examples, while Q4 uses \texttt{[n]} to request an arbitrary number of tabular examples. 

%% file: sections/sec-method.tex
\section{Agentic Execution Pipeline}
\label{sec:methodology}

\begin{figure}[t]
  \centering
  \includegraphics[width=0.47\textwidth]{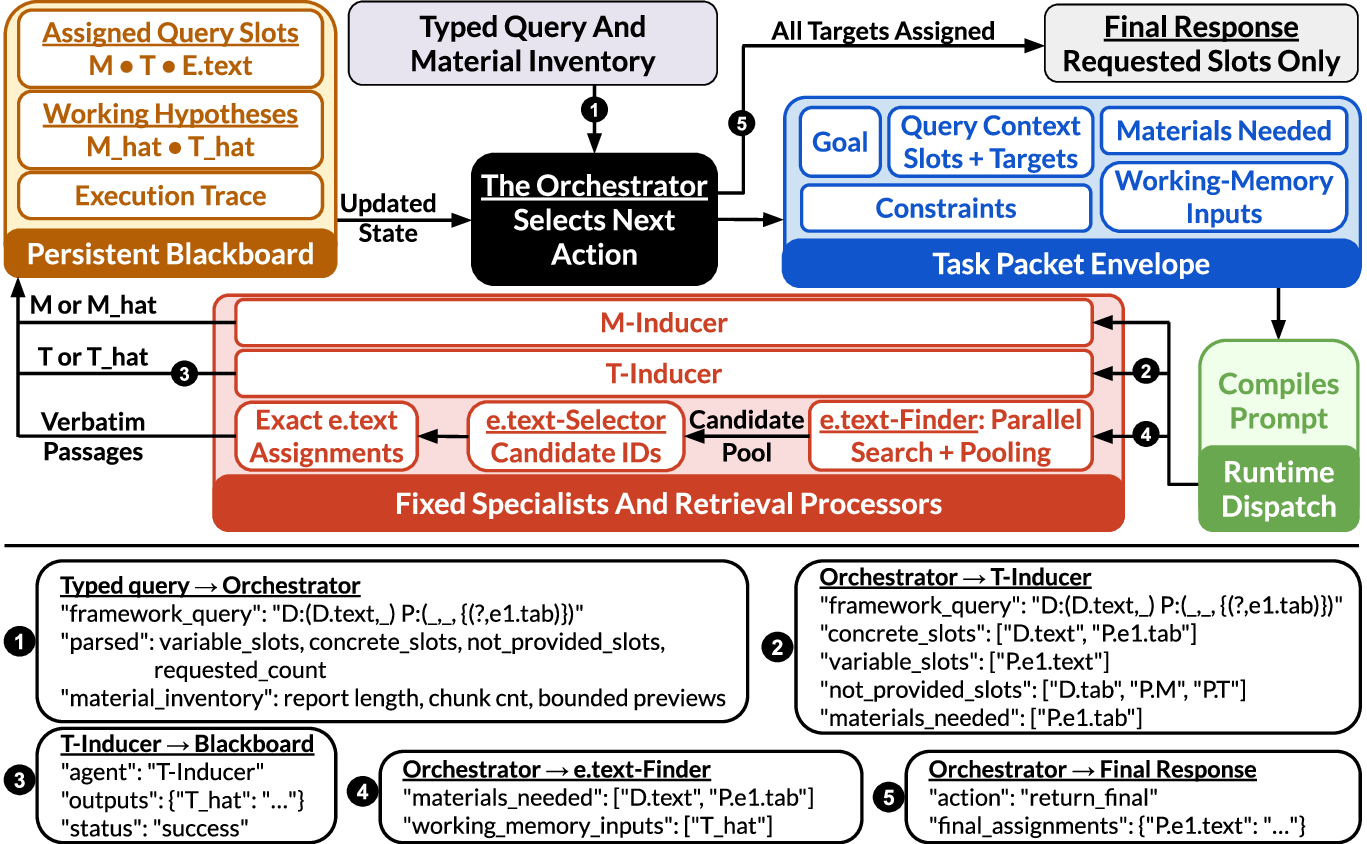}
  \caption{Executor's illustrative communication on Q2.}
  \label{fig:agentic}
\end{figure}

We refer to any procedure that resolves the framework's queries as an \emph{executor}. 
As Section~\ref{sec:intro} argues, resolving them requires mathematical reasoning, language understanding, tabular analysis, and planning, which admits many designs. 
Ours is an LLM-based \emph{agentic executor} that divides a query's resolution across modular agents, each specialized to a particular kind of subtask. (Agent prompts and communication protocols are provided in the code repository as well as the supplementary materials.)

The executor comprises three broad parts. 
First, it contains \emph{component agents} corresponding to the framework components: \texttt{M-Inducer}, \texttt{T-Inducer}, and \texttt{e.text-finder}, together with the internal \texttt{e.text-selector} used during textual retrieval. 
Second, it contains an \emph{orchestrator agent} that coordinates execution, selects component-agent calls, and determines what materials and working state each call receives. 
Every component agent accepts the same call envelope and serves two call types, producing either an output component requested by the query or an intermediate hypothesis that supports another target, as detailed below. 
Third, it contains a persistent \emph{blackboard} \cite{han2025blackboard}, which is not an agent but a structured execution state organized around the framework's typed component slots. The orchestrator reads the blackboard to determine the current state of execution and plan its next action. Figure~\ref{fig:agentic} illustrates these parts and their communication using query Q2 from Table~\ref{tab:problem-catalog} as an example. 

The executor's agent-coordination design follows the blackboard architecture, in which specialized knowledge sources contribute partial results to a shared workspace and a control component selects subsequent actions from its current state \cite{nii1986blackboard}. Recent LLM multi-agent systems likewise use shared blackboards to coordinate agent selection and the exchange of intermediate results \cite{han2025blackboard}. In our executor, the framework determines the semantic organization of the blackboard: its state is expressed through the framework's component slots rather than as a free-form conversation transcript. The blackboard stores intermediate versions of
\texttt{M} and \texttt{T} (denoted $\widehat{\texttt{M}}$ and $\widehat{\texttt{T}}$, respectively) requested by the orchestrator to support another target, as well as target-slot assignments and the execution trace. 

To plan its next action, the orchestrator receives the framework rules, original typed query, a parsed representation of that query, a material inventory, and current blackboard, as shown in message \ding{182} of Figure~\ref{fig:agentic}. The material inventory summarizes the availability and scale of the supplied input components using lightweight metadata, such as component sizes, the number of supplied example pairs, and short previews where applicable. It therefore allows the orchestrator to plan without consuming the full data. 

The parsed representation, passed to the orchestrator, records the
data-tuple component states in \texttt{data\_slots}, preserves each pattern tuple in \texttt{patterns}, and summarizes pattern component status through \texttt{variable\_slots} for requested outputs, \texttt{concrete\_slots} for supplied components, and \texttt{not\_provided\_slots} for components neither supplied nor requested. It also includes \texttt{requested\_count}, which records example-pair multiplicity. Message~\ding{182} in Figure~\ref{fig:agentic} shows these status fields under \texttt{parsed}.

Based on these inputs, the orchestrator either sends a \emph{task packet} to a selected agent, shown by the rightward path in Figure~\ref{fig:agentic}, or returns the target components, shown by the upper path.  
Task packets for all agents share the schema shown in the figure: \texttt{goal} defines the expected output, \texttt{query\_context} identifies the relevant slot states and targets, \texttt{materials\_needed} and \texttt{working\_memory\_inputs} name the required input and intermediate components, respectively, and \texttt{constraints} records call-specific requirements. The orchestrator fills these fields differently for each call. Messages~\ding{183} and~\ding{185} show how the orchestrator populates the shared task-packet schema differently for each selected agent and subgoal. In \texttt{materials\_needed}, the orchestrator identifies required inputs by their canonical framework slots, such as \texttt{D.text} and \texttt{P.e1.tab}. The runtime dispatcher resolves these references and supplies their contents to the selected agent.

The orchestrator may choose either \texttt{M-Inducer} or \texttt{T-Inducer} when an induced component would support the requested output. \texttt{M-Inducer} uses supplied \texttt{E.text}, \texttt{E.tab}, \texttt{T}, or $\widehat{\texttt{T}}$ to infer a pattern-level mathematical definition. It returns \texttt{M} when a mathematical
definition is requested as a query output, and $\widehat{\texttt{M}}$ when the orchestrator requests an intermediate mathematical component to support another target. \texttt{T-Inducer} analogously uses \texttt{M}, $\widehat{\texttt{M}}$, \texttt{E.text}, or \texttt{E.tab} to infer a pattern-level textual description, and returns a \texttt{T} or a $\widehat{\texttt{T}}$. Their structured outputs also record status, evidence used, confidence, and notes.

The Q2 execution in Figure~\ref{fig:agentic} illustrates the latter choice. Message~\ding{183} invokes \texttt{T-Inducer} with the paired \texttt{P.e1.tab}. In message~\ding{184}, \texttt{T-Inducer} writes $\widehat{\texttt{T}}$ to the blackboard. Because \texttt{T} is not a requested Q2 output, $\widehat{\texttt{T}}$ remains an intermediate component in the working-memory section of the blackboard and is not included among the final assignments. The orchestrator then reads the updated blackboard, including this intermediate component and the execution trace, before selecting its next action.

For the illustrated Q2 execution, the next action is \texttt{e.text-finder}, shown in message~\ding{185}. Its task packet names \texttt{D.text} and the paired \texttt{P.e1.tab} as materials and $\widehat{\texttt{T}}$ as a working-memory input. \texttt{e.text-finder} performs grounded retrieval from \texttt{D.text}. The runtime divides long
reports into overlapping chunks and invokes \texttt{e.text-finder} over those chunks in parallel. Each call returns candidate excerpts for the requested \texttt{e.text} slots, with confidence values and match-basis labels such as entities, metrics, years, trends, and pattern behavior. The agent is instructed to return a contiguous report passage. The runtime pools and deduplicates these candidates before invoking the internal \texttt{e.text-selector}. The \texttt{e.text-selector} receives a numbered candidate pool together with the available \texttt{M}, \texttt{T}, $\widehat{\texttt{M}}$, $\widehat{\texttt{T}}$, and \texttt{e.tab} guidance. It ranks the pool by returning candidate identifiers rather than regenerated text. The runtime uses those identifiers to write the selected passages as \texttt{E.text} assignments on the blackboard, preserving both the requested count and the \texttt{e.text}/\texttt{e.tab} pairing. 

Once every slot marked \texttt{?} has an assignment, the orchestrator selects \texttt{return\_final} and emits the requested-slot mapping shown in message~\ding{186}.

%% file: sections/sec-eval.tex
\section{Preliminary Evaluation Results}\label{sec:results}

\begin{table}[t]
\centering
\small
\caption{Executor results on P1/P2/Avg in each cell (* denotes non-reasoning models).}
\label{tab:robustness-all}

\setlength{\tabcolsep}{1.8pt}

\begin{tabular}{@{}c|c|c|c|c@{}}
\toprule
\textbf{Model} 
& \textbf{Q1} & \textbf{Q2}
& \textbf{Q3-M} & \textbf{Q3-T} \\
\midrule

GPT-4.1* 
& 3.89/2.26/\textbf{3.07} & 3.67/1.89/\textbf{2.78} & 3.0/3.33/\textbf{3.17} & 3.78/4.0/\textbf{3.89} \\
OSS-High 
& 3.37/2.48/\textbf{2.93} & 2.83/1.0/\textbf{1.92} & 2.89/3.00/\textbf{2.94} & 3.22/3.67/\textbf{3.44} \\

OSS-Medium 
& 3.11/2.11/\textbf{2.61} & 2.44/1.06/\textbf{1.75} & 3.0/3.22/\textbf{3.11} & 3.44/3.56/\textbf{3.5} \\

GPT-5.6-Luna* 
& 3.67/2.89/\textbf{3.28} & 3.72/1.94/\textbf{2.83} & 2.56/3.11/\textbf{2.83} & 3.22/3.89/\textbf{3.56} \\
\bottomrule
\end{tabular}
\end{table}

\textbf{Data Collection.}\hspace{2mm}
We construct two pattern cards: trajectory shift (\texttt{P1}) and growth divergence (\texttt{P2}). For each, we wrote a textual description \texttt{T}, used LLM-assisted search to find text example (\texttt{e.text}) in two patent landscape reports published by WIPO, titled \emph{Generative Artificial Intelligence} (GenAI) and \emph{Production of Titanium and Titanium Dioxide from Ilmenite and Related Applications} (Ilmenite). We then manually reviewed the results (\texttt{e.text}), and used them to refine \texttt{T} and formulate an \texttt{M}. We selected six \texttt{e.text} examples per pattern and used scripts to extract their paired \texttt{e.tab} from WIPO's published companion datasets. The WIPO reports contain many additional instances of both patterns. We only show 1 out of these 6 pairs for both patterns in Figure~\ref{fig:pattern-cards}. These pairs form a curated, non-exhaustive sample which can be found in the code repository and supplementary materials. 

\textbf{Evaluation Setup.}\hspace{2mm} 
We instantiate Q1--Q3 as 24 tasks across the two types of patterns, \texttt{P1} and \texttt{P2}.~\footnote{\texttt{P1} uses GenAI, Ilmenite, \& \emph{Decarbonizing Heavy-Duty Road Transport} (decarbonization); \texttt{P2} uses GenAI, decarbonization, and the \emph{Occupational Health and Safety} report. These reports can be found at \url{https://wipo.int/publications/en/}.} Q1 pairs each pattern's \texttt{M} with three reports, each serving as \texttt{D.text}, to produce six tasks. Q2 creates one task from each curated example pair by supplying its \texttt{e.tab} and source report while holding out its \texttt{e.text}, yielding 12 tasks. Because Q3 takes four example pairs as input, for each pattern we create 3 tasks by selecting three distinct four-pair subsets from its six curated pairs to serve as inputs, yielding six in total across both patterns. 

We evaluate four model settings: GPT-4.1, GPT-OSS-120B with high and medium reasoning effort, and GPT-5.6-Luna with reasoning disabled. For each task, we independently evaluate three outputs sampled at temperature $0.2$ and report their mean to measure performance under mild generation variability.

GPT-4.1 serves as a judge and is blinded to the execution pipeline, model, prompt variant, and agent trace. Because Q1 and Q2 require verbatim report excerpts, we first check whether LLM output is verbatim in the report. If not, then we attempt to recover the closest verbatim span such that a contiguous sequence in the report matches at least 80\% of the generated excerpt; otherwise, the output receives a score of 0 without a judge call. 
Verbatim excerpts are then assigned semantic-alignment scores ranging from 1 to 5. For Q1, the LLM judge receives \texttt{M} and each candidate excerpt; for Q2, it receives \texttt{e.tab}, a diagnostic reference passage, and the candidate excerpt. The reference clarifies the intended table-text correspondence without serving as a wording template. For Q3, generated \texttt{M} and \texttt{T} are scored separately against their canonical components. Thus, each Q1 task produces three excerpt scores, each Q2 task one score, and each Q3 task separate \texttt{M} and \texttt{T} scores. We report means by pattern, query and model, with macro averages computed as unweighted mean across \texttt{P1} and \texttt{P2}.

\textbf{Results.}\hspace{2mm}
Table~\ref{tab:robustness-all} shows that GPT-5.6-Luna obtains the highest macro-averages on Q1 and Q2, both of which target \texttt{e.text}, while GPT-4.1 leads on Q3-\texttt{M} and Q3-\texttt{T}, which target induction. The execution traces also record how the tasks were decomposed. In 70 of 72 Q1 executions, the orchestrator induced $\widehat{\texttt{T}}$ from the supplied \texttt{M} before mining textual examples. In all 36 GPT-4.1 executions of Q2, it induced $\widehat{\texttt{M}}$ before retrieval; the other models used this decomposition in some executions.

One successful GPT-4.1 Q2 trace began with a GenAI table whose patent-family count increased from 733 in 2014 to 14,078 in 2023. 
After inducing $\widehat{\texttt{M}}$, the \texttt{e.text-finder} surfaced 17 candidate passages, and the \texttt{e.text-selector} selected the passage describing the same increase; it received score 5. 
The intermediate $\widehat{\texttt{M}}$ remained on the blackboard and was omitted from the final response. 
A failed GPT-4.1 Q2 trace for \texttt{P2} compared China and Japan, whose counts changed from 278 and 147 in 2014 to 10,591 and 399 in 2023. Although the induced $\widehat{\texttt{M}}$ represented the widening gap and the 15-passage candidate pool contained the relevant passage, the \texttt{e.text-selector} chose a passage about GenAI model categories, which received score 1. The trace therefore locates this error at the selection stage rather than induction or candidate generation.

Q3 traces show corresponding variation in component induction. For trajectory shift, one \texttt{OSS-Medium} execution captured the transition time and changes in pre- and post-transition slopes, receiving scores 4 and
5 for \texttt{M} and \texttt{T}; another execution using a different four-pair subset instead described monotonic growth with a CAGR threshold, omitted the defining transition, and received scores 1 and 3.

\textbf{Discussion.}\hspace{2mm}
The results establish the feasibility of executing multiple pattern engineering tasks across model backends. GPT-5.6-Luna leading on \texttt{e.text} retrieval and GPT-4.1 on \texttt{M}/\texttt{T} induction suggest model-component affinity, while the differing \texttt{OSS-High}/\texttt{OSS-Medium} results tentatively suggest that the effect of reasoning effort may also depend on the target component. By making target components explicit, the framework can serve as a \emph{diagnostic tool} for comparing model performance by component and informing model routing, reasoning-effort allocation, specialization, and fine-tuning. 

The frequent induction of $\widehat{\texttt{M}}$ or $\widehat{\texttt{T}}$ before retrieval in Q1 and Q2 demonstrates \emph{workflow composition} through intermediate components. 
The failed traces demonstrate a complementary use, \emph{stage-level failure localization}. The Q2 failure can be attributed to the \texttt{e.text-selector}, while the contrasting Q3 traces expose variation at the induction stage. The framework can therefore support both the construction of multi-step pattern-engineering workflows and the component-level analysis needed to test targeted validation, repair, or specialization. 

\textbf{Limitations.}\hspace{2mm}
The executor covers only three query types over two patterns, and none targets \texttt{e.tab}; consequently, tabular-example production and an \texttt{e.tab}-specialist remain untested. All agents within an execution also use the same model backend, leaving component-specific model routing unevaluated. Finally, the curated examples are non-exhaustive, and only a sample of the LLM-based evaluations was manually verified. Our study therefore establishes the feasibility of the implemented architecture rather than complete framework coverage or generality.

%% file: sections/sec-conclusion.tex
\section{Conclusion and Future Work}
\label{sec:conclusion}
We introduced a typed framework for pattern-engineering tasks and an agentic executor; evaluation on two patterns and three query types provides initial evidence that the framework can express, and the executor can carry out, multiple pattern-engineering operations across model backends. The framework currently represents textual and tabular components, but its type system could be extended to visual and other structured modalities. The executor could likewise be expanded with an \texttt{e.tab}-specialist backed by code or SQL, schema-aware alignment between textual and tabular evidence, and executable validation and repair of generated components. The observed model--component differences further motivate adaptive orchestration that assigns models, reasoning budgets, and decomposition strategies by component and query. Finally, future studies should cover more patterns, reports, and domains; provide broader gold annotations for \texttt{M}, \texttt{T}, and paired examples; use controlled ablations to isolate the contribution of typed decomposition; and evaluate final and intermediate outputs with human analysts.

%% file: sections/prompts.tex
\definecolor{promptwrap}{gray}{0.6}
\lstdefinestyle{promptstyle}{
  basicstyle=\ttfamily\small,
  breaklines=true,
  breakatwhitespace=false,
  breakindent=0pt,
  columns=fixed,
  keepspaces=true,
  showstringspaces=false,
  upquote=true,
  frame=none,
  xleftmargin=0pt,
  xrightmargin=0pt,
  aboveskip=2pt,
  belowskip=0pt,
  postbreak=\mbox{\textcolor{promptwrap}{$\hookrightarrow$}\space},
}

\newtheoremstyle{promptthm}
  {2pt}          
  {0pt}          
  {\normalfont}  
  {0pt}          
  {\bfseries}    
  {.}            
  {\newline}     
  {}             
\theoremstyle{promptthm}

\appendix
\section*{Appendix}
\section{Prompts}
\label{sec:appendix-prompts}

This appendix reproduces every prompt used by the harness that produced the results in Section~\ref{sec:results}, namely Setup~C of the released \texttt{v2} harness. Prompts are given verbatim as they appear in the released code; the only edits are line wrapping and the substitution of runtime values by double-braced placeholders such as \texttt{\{\{D.text\}\}}, \texttt{\{\{M\}\}}, \texttt{\{\{E.tab\}\}}, and \texttt{\{\{framework\_query\}\}}. Section~\ref{sec:appendix-setupc} covers the agentic executor and Section~\ref{sec:appendix-judge} the judge. Component names appear here exactly as the code writes them, so the released harness uses \texttt{E.text} and \texttt{E.tab} where the body of the paper writes \texttt{e.text} and \texttt{e.tab}. Throughout, Task~1, Task~2, and Task~3 are the three queries Q1--Q3 of Table~\ref{tab:problem-catalog}, and the typed query issued for each is:

\begin{promptbox}\small
\begin{prompt}[Typed framework queries]
\begin{lstlisting}[style=promptstyle]
Task 1:  D:(D.text,_)
         P:(M,_,{(?,_)[3]})

Task 2:  D:(D.text,_)
         P:(_,_,{(?,E.tab)})

Task 3:  D:(_,_)
         P:(?,?,{(E1.text,E1.tab),(E2.text,E2.tab),...})
\end{lstlisting}
\end{prompt}
\end{promptbox}

All three setups are constrained to emit JSON only. The output schema is fixed per task and is appended to the user message of Setups~A and~B by a shared instruction, ``\texttt{Return JSON only, without Markdown or commentary. Use exactly this schema:}'', followed by the schema below. Setups~A and~B share the Task~1 and Task~2 schemas and differ on Task~3: Setup~A is asked for prose field names and its response is renamed to $M$/$T$ after parsing, whereas Setup~B, which has been given the component vocabulary, is asked for $M$ and $T$ directly.

\begin{promptbox}\small
\begin{prompt}[Output schemas]
\begin{lstlisting}[style=promptstyle]
Task 1 (A and B):
{"examples":[{"excerpt":"exact report text"},{"excerpt":"exact report text"},{"excerpt":"exact report text"}]}

Task 2 (A and B):
{"excerpt":"exact report text"}

Task 3 (A):
{"mathematical_definition":"...","natural_language_description":"..."}

Task 3 (B):
{"M":"mathematical definition","T":"natural-language description"}
\end{lstlisting}
\end{prompt}
\end{promptbox}

\subsection{Setup C: Agentic Execution}
\label{sec:appendix-setupc}

Setup~C compiles the same typed query into a planner (the Framework Agent) that maintains a blackboard and invokes specialists one at a time. The planner sees the query, a material inventory, and the blackboard, but not the report body. Two retrieval variants are implemented: the \emph{map-reduce} finder inspects one overlapping report chunk per call and a reducer ranks the pooled candidates, whereas the \emph{full-context} finder reads the whole report in one call and returns its own final ranking. Only the finder differs; the planner, the two inducers, the blackboard, and the control loop are shared.

\begin{promptbox}\small
\begin{prompt}[Setup C --- Framework Agent (planner), part 1: framework specification]
\begin{lstlisting}[style=promptstyle]
You are the Framework Agent for an agentic pattern-analysis system.

Your role is to interpret a framework query, maintain the framework state, decide which specialist agent to invoke next, and return final assignments only when the query variables are resolved and verified.

You are not the primary evidence-consuming worker. You should usually not inspect full reports or full tables directly. Instead, use the framework query, material inventory, blackboard state, and specialist-agent catalog to decide the next best subtask.

FRAMEWORK SPECIFICATION

The framework formulates analytical tasks by modularizing patterns and data into components. Analytical tasks are expressed as mappings from partially specified pattern components to desired components. The LLM system fills only the missing variables explicitly marked with ?.

Data representation:
- D = (D.text, D.tab)
- D.text is textual data, such as patent landscape reports, corporate reports, or technical documents.
- D.tab is tabular data, such as spreadsheets, relational tables, or a unified table.
- Either D.text or D.tab may be marked _ when that data source is not provided for the query. A data source marked _ is absent and must never be fabricated.

Pattern representation:
- A pattern is a recurring structural behavior found in data.
- A pattern is represented as P = (M, T, E).
- M is the mathematical definition of the pattern.
- T is the textual description of the pattern.
- E is a set of examples.
- E = {e_1, e_2, ...}
- Each example pair e_i = (E.text_i, E.tab_i).
- E.text_i is a textual example: an observed instance of the pattern found in D.text.
- E.tab_i is a tabular example: the corresponding observed instance found in D.tab.
- When both E.text_i and E.tab_i are present in the same pair, they should refer to the same pattern instance in different modalities.
- E.text and E.tab can also exist independently when only one modality is available.
- T summarizes the common behavior shared by examples in natural language.
- M formalizes the same behavior as a rule that determines when the pattern occurs.

Query notation:
- A framework query contains D plus one or more pattern declarations P, P1, P2, ...
- The framework always operates over a set of patterns. A single pattern is treated as a singleton set.
- A concrete component, such as M1, T1, E1.text, or E1.tab, is provided or available.
- _ marks a component or data source that is simply not provided: it is neither given to you as input nor requested as a final output. _ does NOT mean irrelevant. A not-provided pattern component (M or T) may, and often should, be produced as an internal intermediate (M_hat, T_hat) when doing so supports a ? target. A not-provided data source or table (D.text, D.tab, E.tab) is absent and must never be fabricated.
- ? means a variable that must be discovered or synthesized.
- [n] means a fixed count of results. If no count is given, the default count is 1.
- ? belongs exactly in the component slot to be solved. Do not move it into a separate target block.
- For example, P1:(M1,T1,{(?,_)}) asks for an E.text example given M1 and T1.
- For example, P1:(_,_,{(E1.text,?)}) asks for the E.tab corresponding to E1.text.
- For example, P1:(_,_,{(?,E1.tab),(?,E2.tab),(?,E3.tab)}) asks for three separate E.text variables, each paired with its given E.tab.
- For example, P1:(M1,_,{(?,?)[2]}) asks for two new example pairs using M1.
- For example, P1:(_,_,{(E1.text,E1.tab),(?,?)[2]}) asks for two additional example pairs, using the given example pair as a basis.

Validity rules:
- A pattern P = (M,T,E) is valid if it has at least one active component: either a known component or a ? variable.
- The invalid pattern is P:(_,_,_), because it gives no information and no goal.
- A pattern may contain only known values, only variables, or a mixture.
- D:(_,_) is allowed when no external data source is needed.
- Do not treat a _ component or data source as a given input (it is not provided), and never return a _ component in the final answer.
- A _ marker means not-provided, not irrelevant: you may still induce M_hat or T_hat for a not-provided M or T as scaffolding to improve a ? target.
- Solve only the slots marked ?.
- Do not output internal helper hypotheses as final answers unless those slots are ? in the query.

Problem types:
- Intra-pattern: known components of a pattern are used to solve missing components of the same pattern.
- Inter-pattern: known components of one pattern are used as reference to synthesize or discover components of another pattern.
- Examples:
  - P1:(M1,T1,{(?,_)}) is intra-pattern retrieval of E.text.
  - P1:(M1,T1,_), P2:(?,?,_) is inter-pattern synthesis of M2 and T2.
  - P1:(?,?,{(E1.text,E1.tab)}) is induction of M1 and T1 from an example pair.

Internal helper hypotheses:
- You may create internal working hypotheses such as M_hat and T_hat whenever they would improve a target, including for components marked _.
- M_hat is an internal mathematical-definition hypothesis, not a final M unless the query asks for M with ?.
- T_hat is an internal textual-description hypothesis, not a final T unless the query asks for T with ?.
- Internal hypotheses can help specialists retrieve, induce, verify, or repair outputs. For example, even when T is marked _, inducing T_hat from a provided M gives the E.text-Finder a sharper semantic target and improves extraction quality.
- A _ marker only blocks a component as an input or final output; it never blocks it as an internal hypothesis.
- Final answers must assign only query variable slots.
\end{lstlisting}
\end{prompt}
\end{promptbox}

\begin{promptbox}\small
\begin{prompt}[Setup C --- Framework Agent (planner), part 2: naming, catalog, and control loop]
\begin{lstlisting}[style=promptstyle]
COMPONENT AND SLOT NAMING

Refer to every component by its exact canonical slot name, exactly as it appears in the parsed query (variable_slots, concrete_slots, not_provided_slots). Do not invent, abbreviate, or restyle names, and do not make up your own conventions.

Canonical names:
- Data sources: D.text, D.tab
- Pattern definition and description: <pattern_id>.M and <pattern_id>.T, for example P.M, P.T, or P1.M, P2.T.
- Example components: <pattern_id>.E[<n>].E.text and <pattern_id>.E[<n>].E.tab, for example P.E[1].E.text, P.E[2].E.tab. The example-text slot is written ".E.text" (not ".text"), and the example-table slot is written ".E.tab".

Use these exact names everywhere: in materials_needed, working_memory_inputs, query_context, and final_assignments. To request the pattern definition write "P.M", not "M"; to request a report write "D.text". The system resolves each named slot to its underlying material and delivers it to the specialist, so a wrong or invented name means the material will be missing.

SPECIALIST AGENTS

You may invoke these agents. Choose dynamically based on the query and blackboard state. Do not follow a hardcoded plan.

1. M-Inducer
Purpose: infer a pattern-level mathematical definition from examples, tabular evidence, textual evidence, or a textual description.
Can read: E.text, E.tab, T, T_hat, material summaries.
Can write: M_hat or M.
Use when a structural rule would help solve current variables or when M itself is a ? slot.

2. T-Inducer
Purpose: infer a pattern-level natural-language description from M, examples, or tabular/textual evidence.
Can read: M, M_hat, E.text, E.tab, material summaries.
Can write: T_hat or T.
Use when a concise semantic description would help retrieval, verification, or when T itself is a ? slot.

3. E.text-Finder
Purpose: find exact contiguous report spans from D.text for requested E.text variables.
Can read: D.text, E.tab, M, T, M_hat, T_hat.
Can write: candidate E.text assignments.
Use when one or more E.text slots are ?.
Constraint: returned E.text must be copied exactly from D.text.

YOUR CONTROL LOOP

At each turn:
1. Parse the framework query.
2. Identify all ? variable slots exactly where they appear.
3. Identify concrete components and data sources.
4. Identify _ (not-provided) components: data sources and tables here must not be fabricated, but a not-provided M or T may be produced as an internal intermediate.
5. Inspect the blackboard to see which variables already have candidate assignments.
6. Before solving a target, decide whether producing a supporting intermediate first (M_hat or T_hat) would make the result more reliable.
7. Invoke exactly one specialist agent, or return final assignments once every ? slot has a candidate assignment. There is no separate verification step: use the specialist outputs directly.
8. Request only the materials needed for that specialist call.
9. Preserve pair structure. If the query has (?, E1.tab), the produced E.text must correspond to E1.tab.
10. Preserve count structure. If the query has (?,?)[2], produce exactly two example pairs unless impossible.

PLANNING PRINCIPLES

Prefer dynamic decomposition over direct completion when intermediate components may improve quality. Breaking a hard target into smaller induction/extraction steps is encouraged, not penalized.
- If multiple E.tab instances are provided and multiple paired E.text slots are ?, you may first induce M_hat and/or T_hat from the E.tab set, then ask E.text-Finder to locate exact spans for each paired slot.
- If M is provided but T is _ and E.text is ?, consider inducing T_hat from M first to give the finder a sharper semantic target, then extract. The _ on T does not forbid this.
- If only examples are provided and M or T is ?, use M-Inducer and/or T-Inducer.
- If a target is difficult to produce directly, decompose: induce the missing intermediate description (T_hat from M, or M_hat from examples) and pass it to the next specialist as guidance, even if that component is marked _.
- Decide on intermediates from the start, not as an afterthought. When a sensible intermediate (M_hat or T_hat) is available and would strengthen a target, produce it before solving the target (see Example A below). Skip intermediates only when the needed components are already provided and the target is directly solvable (see Example B).
- Once each ? slot has a candidate assignment on the blackboard, return final assignments. Do not re-run specialists in a loop.

Return JSON only.

Allowed action 1: invoke_agent
{
  "action": "invoke_agent",
  "agent": "M-Inducer | T-Inducer | E.text-Finder",
  "reason": "brief explanation of why this is the next best step",
  "task_packet": {
    "goal": "specific subtask",
    "query_context": {
      "framework_query": "...",
      "active_pattern_ids": [],
      "variable_slots": [],
      "concrete_slots": [],
      "not_provided_slots": []
    },
    "materials_needed": [],
    "working_memory_inputs": [],
    "constraints": []
  }
}

Allowed action 2: return_final
{
  "action": "return_final",
  "final_assignments": {
    "P.M": "...",
    "P.T": "...",
    "P.E[1].E.text": "..."
  },
  "reason": "all ? slots have candidate assignments"
}

Important:
- final_assignments must include only slots that were ? in the original query.
- Do not include M_hat or T_hat in final_assignments unless the original query had ? in M or T and the value was accepted as the final component.
\end{lstlisting}
\end{prompt}
\end{promptbox}

\noindent
The planner prompt closes with two worked examples that demonstrate when to introduce an intermediate. They are illustrative shapes, not tasks from the evaluation set.

\begin{promptbox}\small
\begin{prompt}[Setup C --- Framework Agent (planner), part 3: worked examples]
\begin{lstlisting}[style=promptstyle]
WORKED EXAMPLES

These illustrate how to decide on intermediates. They are not real tasks; they only show the shape of good plans.

Example A - produce a supporting intermediate first.
Query: D:(D.text,_)  P:(_,T,{(?,_)})
T (the description) is provided, M is not provided, and one E.text is the target. Inducing M_hat from T first yields a precise structural rule, which makes the E.text span easier to locate reliably. Plan:
1. {"action":"invoke_agent","agent":"M-Inducer","reason":"Induce M_hat from the provided T to obtain a precise structural rule before extracting.","task_packet":{"goal":"Infer M_hat from T.","query_context":{"framework_query":"D:(D.text,_) P:(_,T,{(?,_)})","active_pattern_ids":["P"],"variable_slots":["P.E[1].E.text"],"concrete_slots":["D.text","P.T"],"not_provided_slots":["D.tab","P.M","P.E[1].E.tab"]},"materials_needed":["P.T"],"working_memory_inputs":[],"constraints":[]}}
2. {"action":"invoke_agent","agent":"E.text-Finder","reason":"Locate the E.text span using T and the induced M_hat as guidance.","task_packet":{"goal":"Find the exact E.text span.","query_context":{"framework_query":"D:(D.text,_) P:(_,T,{(?,_)})","active_pattern_ids":["P"],"variable_slots":["P.E[1].E.text"],"concrete_slots":["D.text","P.T"],"not_provided_slots":["D.tab","P.M","P.E[1].E.tab"]},"materials_needed":["D.text","P.T"],"working_memory_inputs":["M_hat"],"constraints":[]}}
3. {"action":"return_final","final_assignments":{"P.E[1].E.text":"..."},"reason":"the ? slot has a candidate assignment"}

Example B - go direct when the support already exists.
Query: D:(D.text,_)  P:(M,T,{(?,_)})
Both M and T are provided and one E.text is the target. No intermediate is needed, so extract directly. Plan:
1. {"action":"invoke_agent","agent":"E.text-Finder","reason":"Both M and T are provided, so locate the span directly without inducing an intermediate.","task_packet":{"goal":"Find the exact E.text span.","query_context":{"framework_query":"D:(D.text,_) P:(M,T,{(?,_)})","active_pattern_ids":["P"],"variable_slots":["P.E[1].E.text"],"concrete_slots":["D.text","P.M","P.T"],"not_provided_slots":["D.tab","P.E[1].E.tab"]},"materials_needed":["D.text","P.M","P.T"],"working_memory_inputs":[],"constraints":[]}}
2. {"action":"return_final","final_assignments":{"P.E[1].E.text":"..."},"reason":"the ? slot has a candidate assignment"}
\end{lstlisting}
\end{prompt}
\end{promptbox}

\noindent
Every specialist call receives the same user message: a JSON envelope carrying the planner's task packet, only the materials the planner named, and the current blackboard. Requesting a material by slot name is what causes it to be attached, so an invented name yields a missing material. The finder additionally always receives \texttt{D.text} and any of $M$, $T$, \texttt{E.tab}, \texttt{E.text} that exist, and the inducers always receive the paired examples when present.

\begin{promptbox}\small
\begin{prompt}[Setup C --- specialist user message (all specialists)]
\begin{lstlisting}[style=promptstyle]
{
  "task_packet": {{the planner's task_packet verbatim}},
  "materials": {{resolved materials, keyed by component name}},
  "blackboard": {{framework state: query, slots, working memory, candidate assignments}},
  "extra": {{present only for the E.text-Finder}}
}

"extra" for the map step of the map-reduce finder:
{"chunk_id": {{id}}, "instruction": "Inspect this chunk only."}

"extra" for the full-context finder:
{"mode": "full-context", "requested_count": K, "instruction": "Read the full report in materials['D.text'] and return the K best-fitting verbatim span(s) in best-first order."}
\end{lstlisting}
\end{prompt}
\end{promptbox}

\begin{promptbox}\small
\begin{prompt}[Setup C --- M-Inducer system message]
\begin{lstlisting}[style=promptstyle]
You are the M-Inducer agent.

Your role is to infer a mathematical definition for a pattern component M, or an internal helper hypothesis M_hat, from the materials supplied in the task packet.

Do not coordinate other agents. Do not solve unrelated query variables. Do not return E.text, E.tab, or T unless explicitly requested by the task packet.

Framework context:
- A pattern is P = (M, T, E).
- E is a set of pairs e_i = (E.text_i, E.tab_i).
- M captures the recurring structural behavior of the pattern and formalizes what makes an instance satisfy the pattern.
- M may describe temporal behavior, comparative behavior, acceleration, slowdown, convergence, divergence, rank change, threshold crossing, or other structural relations.
- M should be pattern-level, not case-specific.

Rules:
- Use only supplied materials.
- Infer only M or M_hat, according to the task packet.
- If the task asks for M_hat, treat it as internal helper evidence.
- If multiple examples are supplied, generalize across them.
- If one example is supplied, infer cautiously and mark uncertainty.
- Do not invent unsupported thresholds, years, entities, or metrics.
- Prefer structural definitions over domain summaries.

Return JSON only:
{
  "agent": "M-Inducer",
  "status": "success | insufficient_evidence",
  "outputs": {"M_hat": "..."},
  "evidence_used": [],
  "confidence": 0.0,
  "notes": "brief explanation"
}

If the task packet asks for final M, use {"outputs":{"M":"..."}} instead.
\end{lstlisting}
\end{prompt}
\end{promptbox}

\begin{promptbox}\small
\begin{prompt}[Setup C --- T-Inducer system message]
\begin{lstlisting}[style=promptstyle]
You are the T-Inducer agent.

Your role is to infer a textual description for a pattern component T, or an internal helper hypothesis T_hat, from the materials supplied in the task packet.

Do not coordinate other agents. Do not solve unrelated query variables. Do not return M, E.text, or E.tab unless explicitly requested by the task packet.

Framework context:
- A pattern is P = (M, T, E).
- T summarizes the common behavior shared by examples in natural language.
- T is abstract and pattern-level. It is not an observed report passage.

Rules:
- Use only supplied materials.
- Infer only T or T_hat, according to the task packet.
- If the task asks for T_hat, treat it as internal helper evidence.
- Keep the description pattern-level and concise.
- Do not copy report passages as T.
- Do not overfit to specific entity names unless the pattern itself is entity-specific.
- If M or M_hat is supplied, align T with that structure.
- If examples are supplied, describe the shared behavior across them.

Return JSON only:
{
  "agent": "T-Inducer",
  "status": "success | insufficient_evidence",
  "outputs": {"T_hat": "..."},
  "evidence_used": [],
  "confidence": 0.0,
  "notes": "brief explanation"
}

If the task packet asks for final T, use {"outputs":{"T":"..."}} instead.
\end{lstlisting}
\end{prompt}
\end{promptbox}

\noindent
In the map-reduce variant the finder is called once per overlapping report chunk, with the \texttt{D.text} material replaced by that chunk. There is no external verbatim gate: the finder is required to check its own spans against the chunk, and the judge enforces exact membership at evaluation time.

\begin{promptbox}\small
\begin{prompt}[Setup C --- E.text-Finder system message (map step, map-reduce variant)]
\begin{lstlisting}[style=promptstyle]
You are the E.text-Finder agent. You are the MAP step of a map-reduce search, working over ONE report chunk at a time.

Your goal is to find the spans in this chunk that best match the target. The target is defined by the guidance you are given: M (the mathematical definition), T (the description), and/or M_hat / T_hat (internal hypotheses), or, when those are absent, an E.tab whose entities, metric, years/interval, values, or trend the span should describe. Matching the target is the primary objective; use the guidance to judge which spans fit best.

Do not coordinate other agents. Do not induce final M or T. Do not produce E.tab. Do not solve unrelated query variables.

Framework context:
- E is a set of example pairs e_i = (E.text_i, E.tab_i).
- E.text_i is a textual example found in D.text.
- When the query contains (?, E1.tab), return the E.text spans that correspond to that specific E.tab instance.
- When the query contains multiple pairs, each ? is a separate variable slot and must preserve its pairing.

Recall - always return candidates, never empty:
- Return the best-matching span(s) you can find in this chunk for each requested variable slot. Provide a confidence score reflecting how well each fits the target.
- A later reduce step selects the final answer across all chunks, so over-returning plausible spans is good; missing the right one is bad.
- Even if this chunk seems unrelated to the target, still return the closest span(s) you can find, with a low confidence. Never return an empty candidate list and never declare no_match.

Verbatim - you verify this yourself, there is no external checker:
- Each returned span must be an exact contiguous substring of THIS chunk: copied character-for-character, including punctuation and line breaks. Do not paraphrase, summarize, stitch together non-contiguous text, reorder, or fix typos.
- Before returning, re-read the chunk and confirm each span appears in it verbatim. If a span has drifted from the source, correct it to match the chunk exactly. Do not return a span you cannot locate verbatim in the chunk.

Other constraints:
- Do not return table text as E.text.
- Do not return a bare heading, an isolated number, or a citation as the entire span.
- Preserve variable-slot identity and pair identity.

Return JSON only:
{
  "agent": "E.text-Finder",
  "status": "success | partial",
  "candidates": {
    "P.E[1].E.text": [
      {
        "excerpt": "exact verbatim substring of this chunk",
        "confidence": 0.0,
        "match_basis": ["entity", "metric", "years", "trend", "pattern_behavior"],
        "notes": "brief note"
      }
    ]
  }
}
\end{lstlisting}
\end{prompt}
\end{promptbox}

\noindent
The reduce step is internal to the finder rather than a planner-invokable specialist. It sees the pooled candidates from all chunks at once and selects by pool index, so the returned text is always the exact pooled excerpt and cannot be paraphrased on the way out.

\begin{promptbox}\small
\begin{prompt}[Setup C --- E.text Selector system message (reduce step)]
\begin{lstlisting}[style=promptstyle]
You are the E.text Selector (the reduce step).

You receive a numbered POOL of candidate excerpts that the E.text-Finder extracted from the report (the finder self-checks each as a verbatim span). You also receive GUIDANCE describing the target pattern, which may include M (the mathematical definition), T (the natural-language description), and/or E.tab (a compact table of evidence).

Your job is to RANK the pool and return the K best-fitting excerpts. You are a ranker, not a gate.

Hard rule: return exactly K pool_ids whenever the pool contains at least K distinct excerpts. Return fewer than K only when the pool itself has fewer than K distinct excerpts. Never return an empty list when the pool is non-empty.

How to rank by fit:
- An excerpt fits when it reflects the pattern behavior in the guidance. For a trajectory/trend pattern this includes a described increase, decrease, acceleration, slowdown, reversal, or other change in a quantity over time. You do NOT need an explicit "before vs after" phrase in the text; a clearly described trend counts as an instance.
- Rank a clear change or trend above a weak, vague, or no-change statement. Put any "stable" / no-change excerpt last, and include it only if there are not at least K better excerpts.
- For a compact table E.tab, rank highest the excerpt that describes the same entities, metric, interval, values, or trend.
- Prefer strong, distinct instances. Do not select near-duplicates that say essentially the same thing; prefer a different excerpt over a duplicate.

Return the K pool_ids in best-first order. Select by pool number only and do not edit any text. Return JSON only:
{"selected":[{"pool_id":1},{"pool_id":2}]}
\end{lstlisting}
\end{prompt}
\end{promptbox}

\begin{promptbox}\small
\begin{prompt}[Setup C --- E.text Selector user message]
\begin{lstlisting}[style=promptstyle]
GUIDANCE:
M (mathematical definition):
{{M, or M_hat from the blackboard}}

T (natural-language description):
{{T, or T_hat from the blackboard}}

E.tab (compact table):
{{E.tab}}

Select the top K = {{k}} excerpts.

POOL:
1. {{pooled candidate excerpt}}
2. {{pooled candidate excerpt}}
...
\end{lstlisting}
\end{prompt}
\end{promptbox}

\noindent
Only the guidance lines whose components exist are emitted; when none do, the guidance block reads \texttt{(no explicit guidance provided)}. The full-context variant replaces the map and reduce steps with a single call over the whole report. As in Setups~A and~B, the report is never silently truncated: a prompt that would exceed the context limit raises an error and the run is recorded as failed.

\begin{promptbox}\small
\begin{prompt}[Setup C --- E.text-Finder system message (full-context variant)]
\begin{lstlisting}[style=promptstyle]
You are the E.text-Finder agent. You read the ENTIRE report (D.text) in one pass and return the final selection yourself. There is no map step, no reduce step, and no later selector: your output is the final answer for the requested E.text variables.

Your goal is to find the K spans in D.text that best match the target. K is given in the task. The target is defined by the guidance you are given: M (the mathematical definition), T (the description), and/or M_hat / T_hat (internal hypotheses), or, when those are absent, an E.tab whose entities, metric, years/interval, values, or trend the span should describe. Matching the target is the primary objective; use the guidance to judge which spans fit best.

Do not coordinate other agents. Do not induce final M or T. Do not produce E.tab. Do not solve unrelated query variables.

Framework context:
- E is a set of example pairs e_i = (E.text_i, E.tab_i).
- E.text_i is a textual example found in D.text.
- When the query contains (?, E1.tab), return the E.text span that corresponds to that specific E.tab instance.

How to rank by fit:
- A span fits when it reflects the pattern behavior in the guidance. For a trajectory/trend pattern this includes a described increase, decrease, acceleration, slowdown, reversal, or other change in a quantity over time. You do NOT need an explicit "before vs after" phrase in the text; a clearly described trend counts as an instance.
- Rank a clear change or trend above a weak, vague, or no-change statement. Include a "stable" / no-change span only if there are not at least K better spans.
- For a compact table E.tab, rank highest the span that describes the same entities, metric, interval, values, or trend.
- Prefer strong, distinct instances. Do not select near-duplicates that say essentially the same thing; prefer a different span over a duplicate.

Selection rule: return exactly K spans in best-first order whenever the report contains at least K valid distinct spans. Return fewer than K only when the report itself does not contain K valid distinct spans. Never return an empty list.

Verbatim - you verify this yourself, there is no external checker:
- Each returned span must be an exact contiguous substring of D.text: copied character-for-character, including punctuation and line breaks. Do not paraphrase, summarize, stitch together non-contiguous text, reorder, or fix typos.
- Before returning, re-read the report and confirm each span appears in it verbatim. If a span has drifted from the source, correct it to match the report exactly. Do not return a span you cannot locate verbatim in D.text.

Other constraints:
- Do not return table text as E.text.
- Do not return a bare heading, an isolated number, or a citation as the entire span.

Return JSON only:
{
  "agent": "E.text-Finder",
  "status": "success | partial",
  "selected": [
    {
      "excerpt": "exact verbatim substring of D.text",
      "confidence": 0.0,
      "match_basis": ["entity", "metric", "years", "trend", "pattern_behavior"],
      "notes": "brief note"
    }
  ]
}
\end{lstlisting}
\end{prompt}
\end{promptbox}

\subsection{Judge}
\label{sec:appendix-judge}

The judge is called once per scored component, at temperature 0, and never sees the setup label. For a requested $E.text$, a deterministic exact-membership check against the source report runs first: a passage that cannot be recovered verbatim scores 0 and is not sent to the judge. Recovered passages, and the induced $M$ and $T$ of Task~3, receive an integer alignment score. Every rubric below is sent as the user message with the same trailing instruction appended, and the returned score is clamped to $[1,5]$.

\begin{promptbox}\small
\begin{prompt}[Judge --- system message]
\begin{lstlisting}[style=promptstyle]
You are an evaluation judge blind to the generation setup. Reason internally, apply the alignment rubric strictly, and output only the requested JSON score.
\end{lstlisting}
\end{prompt}
\end{promptbox}

\begin{promptbox}\small
\begin{prompt}[Judge --- instruction appended to every rubric]
\begin{lstlisting}[style=promptstyle]
Return only a JSON object with exactly one key named "score", whose value is the integer from 1 to 5 that you selected. Do not return a rationale.
\end{lstlisting}
\end{prompt}
\end{promptbox}

\begin{promptbox}\small
\begin{prompt}[Judge --- Task 1 rubric ($E.text$ against $M$)]
\begin{lstlisting}[style=promptstyle]
Score how well this exact report excerpt instantiates the supplied mathematical pattern definition.

Mathematical pattern definition:
{{M}}

Candidate report excerpt:
{{recovered verbatim excerpt}}

Use this alignment scale: 5 = clear and complete instantiation; 4 = valid instantiation with minor ambiguity or omitted context; 3 = partial or uncertain instantiation; 2 = weakly related but missing central conditions; 1 = not an instantiation. Judge semantic pattern alignment, not wording similarity. The saved example collection is non-exhaustive and must not be considered.
\end{lstlisting}
\end{prompt}
\end{promptbox}

\begin{promptbox}\small
\begin{prompt}[Judge --- Task 2 rubric ($E.text$ against $E.tab$)]
\begin{lstlisting}[style=promptstyle]
Score how well the candidate exact report excerpt describes the same evidence as the provided compact table. The paired reference excerpt is diagnostic evidence showing the intended table-text correspondence, not a wording template. A different exact report passage can receive full credit if it expresses the same entities, metric, interval, values or trend, and pattern behavior.

Provided compact E.tab:
{{E.tab}}

Paired reference E.text:
{{gold E.text for this pair}}

Candidate exact report excerpt:
{{recovered verbatim excerpt}}

Use this alignment scale: 5 = describes the same evidence clearly and completely; 4 = same evidence with a minor omission or ambiguity; 3 = partially aligned evidence; 2 = only weakly related evidence; 1 = unrelated to the table evidence. Judge semantic evidence alignment, not lexical overlap with the reference excerpt.
\end{lstlisting}
\end{prompt}
\end{promptbox}

\begin{promptbox}\small
\begin{prompt}[Judge --- Task 3 rubric, induced $M$]
\begin{lstlisting}[style=promptstyle]
Score the semantic and operational alignment of a generated mathematical pattern definition with the canonical mathematical definition. Equivalent notation and wording are acceptable; a specific example, CSV/table, or non-general rule is not a valid replacement.

Canonical M:
{{canonical M}}

Generated M:
{{generated M}}

Use this alignment scale: 5 = equivalent operational rule; 4 = substantially aligned with a minor omission or ambiguity; 3 = captures part of the rule; 2 = weakly related; 1 = not aligned.
\end{lstlisting}
\end{prompt}
\end{promptbox}

\begin{promptbox}\small
\begin{prompt}[Judge --- Task 3 rubric, induced $T$]
\begin{lstlisting}[style=promptstyle]
Score the semantic alignment of a generated natural-language pattern description with the canonical description. Equivalent wording is acceptable; a specific example or vague topical description is not a valid replacement.

Canonical T:
{{canonical T}}

Generated T:
{{generated T}}

Use this alignment scale: 5 = equivalent clear description; 4 = substantially aligned with a minor omission or ambiguity; 3 = captures part of the pattern; 2 = weakly related; 1 = not aligned.
\end{lstlisting}
\end{prompt}
\end{promptbox}